\documentclass[aps,twocolumn,pra,superscript,floatfix,superscriptaddress,showpacs,footinbib]{revtex4-1}

\usepackage{natbib}
\usepackage{amssymb}

\usepackage[pdftex]{graphicx}
\usepackage{dcolumn}
\usepackage{bm}
\usepackage{amsmath}
\usepackage{array}
\usepackage{color}
\usepackage{float}
\usepackage{subfigure}
\usepackage{dsfont}
\usepackage{txfonts}
\usepackage{wasysym}
\usepackage{multirow}
\usepackage{sidecap}
\usepackage{xcolor,cancel}
\usepackage[normalem]{ulem}
\usepackage{natmove}  

\newcommand{\dn}{\downarrow}

\newcommand{\e}{\varepsilon}

\newcommand{\up}{\uparrow}
\newcommand{\down}{\downarrow}

\newcommand{\veck}{\mathbf{k}}

\usepackage{hyperref}
\hypersetup{
    colorlinks,%
    citecolor=blue,%
    linkcolor=blue,%
    urlcolor=blue
}

\begin{document}

\title{Kondo screening regimes in multi-Dirac and Weyl systems }

\author{G. T. D. Pedrosa}
\affiliation{Instituto de F\'isica, Universidade Federal de Uberl\^andia, 
Uberl\^andia, Minas Gerais 38400-902, Brazil.}

\author{Joelson F. Silva}
\affiliation{Instituto de F\'isica, Universidade Federal de Uberl\^andia, 
Uberl\^andia, Minas Gerais 38400-902, Brazil.}

\author{E. Vernek}
\affiliation{Instituto de F\'isica, Universidade Federal de Uberl\^andia, 
Uberl\^andia, Minas Gerais 38400-902, Brazil.}
%

\date{\today}

\begin{abstract}
We have investigated the Kondo physics of a single magnetic impurity embedded in multi-Dirac (Weyl) node fermionic systems. By using a generic effective model for the host material and employing a numerical renormalization group approach we access the low temperature behavior of the system, identifying the existence of Kondo screening in single-, double-, and triple-Dirac (Weyl) node models. We find that in any multi-Dirac node systems the low-energy regime lies within one of the known classes of pseudogap Kondo problem, extensively studied in the literature. Kondo screening is also observed  for time reversal symmetry broken Weyl systems. This is, however, possible only in the particle-hole symmetry broken regime obtained for finite chemical potential $\mu$. Although weakly, breaking time-reversal symmetry suppresses the Kondo resonance, especially in the single-node Weyl semimetals. More interesting Kondo screening regimes are obtained for inversion symmetry broken multi-Weyl fermions.  In these systems the Kondo regimes of  double- and triple-Weyl node models are much richer than in the single-Weyl node model. While in the single-Weyl node model the Kondo temperature increases monotonically with  $|\mu|$ regardless the value of the inversion symmetry breaking parameter $Q_0$, in double- and triple-Weyl node models there are two distinct regimes: (i) For $Q_0< |\mu|$ the Kondo temperature depends strongly on $\mu$, while (ii) for $Q_0 > |\mu|$ the Kondo temperature depends very weakly on $\mu$, resembling the flat-band single impurity Anderson model.        

\end{abstract}
\maketitle

\section{Introduction}
\label{sec:Introduction}
\noindent
Topological materials have promoted a colossal excitement in recent years~\cite{doi:10.7566/JPSJ.82.102001, PhysRevLett.95.226801, Bernevig1757,PhysRevB.76.045302, Xia2009}. The  idea behind the characterization  of these systems is the use of concepts of topology to classify their electronic band structures. As such, topological phase transition involves changes in some topological index rather than in an  order parameter by a  symmetry break~\cite{PhysRevB.84.235126,PhysRevX.6.021008}. The best known example of topological materials are topological insulators~\cite{10.2307/j.ctt19cc2gc,RevModPhys.82.3045,doi:10.7566/JPSJ.82.102001} that are bulk insulators but exhibit band touching (metallic) edge states that are protected by time reversal symmetry (TRS)\cite{RevModPhys.82.3045,Moore2010, Konig766,RevModPhys.83.1057}. Analogs of these 2D materials exists in 3D and are generically dubbed Dirac semimetals~\cite{PhysRevLett.108.140405} found in compounds such as Na$_3$Bi~\cite{Liu864} and Cd$_3$As$_2$~\cite{PhysRevLett.113.027603, PhysRevLett.113.246402}. Near the crossing points, the linearly dispersive bands are well described by 3D versions of the Dirac equation. These crossing points are therefore commonly called Dirac nodes, and their degeneracy is protected by both TRS and inversion symmetry (IS) of the lattice structure.

An interesting family of topological materials called Weyl semimetals (WSM) emerge from Dirac systems when at least one of the aforementioned  symmetries that protect the degeneracy of the Dirac nodes is  broken~\cite{RevModPhys.90.015001,Zhong2016,PhysRevLett.108.266802, Yang2014}. These systems are recognized by exhibiting peculiar physical properties such as Fermi arcs \cite{PhysRevB.83.205101, Xu613,PhysRevB.87.245112}, chiral anomaly~\cite{PhysRevLett.119.176804,PhysRevB.101.165309}, affecting drastically their transport properties~\cite{Kundu_2020, PhysRevB.97.125419}. 
Another variety of WSMs was predicted to exist protected by group symmetry C$_n$~\cite{PhysRevLett.108.266802}. These WSMs exhibit topological charge $J>1$ and appear as a generalization of the first WSM exhibiting $J=1$, as such they were called multi-Weyl node semimetals~(MWSMs) \cite{PhysRevLett.108.266802}. This prediction was confirmed for $J=2$ in HgCr$_2$Se$_4$ and SrSi$_2$~\cite{PhysRevLett.107.186806, PhysRevLett.108.266802,Huang1180} while systems in which $J=3$ have been anticipated to exist in quasi 1D molybdenum monochalcogenide compounds~\cite{PhysRevX.7.021019}.

With the increasing popularity of these exotic materials and the gain of comprehension on their properties, we have witnessed a rising interest in the low-temperature properties of these materials when they host magnetic impurities~\cite{PhysRevB.92.041107,yanagisawa}. When an isolated quantum magnetic impurity (also known as Anderson impurity) is inserted in a system of free electrons, such as metals and also these topological materials, their itinerant electrons cooperate to screen the localized moment, a phenomena known as the Kondo effect~\cite{PhysRev.124.41}. Although the Kondo effect is more commonly studied in conventional  metallic materials \cite{HewsonBook,Bulla}, interesting facets of Kondo physics is also found in structured  conduction bands near the Fermi level~\cite{PhysRevLett.97.096603, PhysRevLett.119.116801,PhysRevB.57.14254,PhysRevB.90.075150}. More recently, Kondo effect has also been investigated in topological materials such as Dirac and Weyl semimetals in which metallic or pseudogap screening regimes have been found \cite{PhysRevB.92.041107, yanagisawa,PhysRevB.81.241414}. 

In this work we revisit the problem of Kondo screening regimes in the Dirac and Weyl systems~\cite{Mitchell}, with special attention to the MWSMs, on which a detailed investigation of the Kondo effect is still lacking. We adopt a generic model capable of describing multi-Dirac semimetals~(MDSMs) as well as MWSMs~\cite{PhysRevB.99.115109} and employ a numerical renormalization group (NRG) approach~\cite{HewsonBook,Bulla} to access the Kondo physics of the system in a systematic manner. Our numerical results show that the Kondo regimes of all MDSM lie on some class of pseudogap Kondo effect, in which Kondo screening is possible only at the particle-hole asymmetric situation. In contrast, in IS broken  MWSM, Kondo screening is possible even in particle-hole symmetric condition. Moreover, rich Kondo regimes emerge as a result of the interplay between the chemical potential $\mu$ and  inversion symmetry parameter $Q_0$.  We also find that in  MWSMs with TRS broken the Kondo screening is affected, albeit weakly, by magnetic polarization induced at the impurity via hybridization function. 

The rest this paper is organized as follows. In Sec.~\ref{model} we present the quantum impurity model Hamiltonian to describe the physical system and discuss how to approach its Kondo regimes. In Sec.~\ref{numerical_results}  we present our numerical results and discussion the various regimes of the model and, finally, in Sec.~\ref{conclusions} we  bring our concluding remarks.
 
\section{Model and Method}
\label{model}


For completeness, the total Hamiltonian that describes our system can be split into three terms as ${H} = {H}_{0}+{H}_{\rm imp}+{H}_{\rm V}$, where ${H}_{0}$ represents the clean multi-Weyl/Dirac fermion system, ${H}_{\rm imp}$ describes the impurity Hamiltonian, and ${H}_{\rm V}$ 
accounts for the hybridization between them. Introducing the spinor $\Psi_{\textbf{k}}^{\dagger}= (c_{\mathbf{k+\uparrow}}^{\dagger} ,c_{\mathbf{k+\downarrow}}^{\dagger} ,c_{\mathbf{k-\uparrow}}^{\dagger}, c_{\mathbf{k-\downarrow}}^{\dagger})$, in which  $c^\dagger_{\veck ps}$ creates a fermion with momentum $\veck$, chirality $p=+,-$ and spin $s=\up, \dn$, we can write ${H}_{0} = \sum_{\textbf{k}}\Psi_{\textbf{k}}^{\dagger}{\cal H}_{\textbf{k}} \Psi_{\textbf{k}}$
%
%
with
\begin{eqnarray}
{\cal H}_{\mathbf{k}} &=& \tau_{z}\otimes\Bigg[ v_{\perp}k_{0}\Big( \tilde{k}^{J}_{-}\sigma_{+} + \tilde{k}^{J}_{+}\sigma_{-}\Big) +  v_{z}k_{z}\sigma_{z}-Q_{0}\sigma_{0}\Bigg] \notag \\
&& -v_{z}Q\tau_{0} \otimes\sigma_{z}-\mu \tau_0\otimes\sigma_0.
\label{hamiltonian:model}
\end{eqnarray}
%
%
Here,  $\mu$ is the chemical potential, $v_{\perp}$ and $v_{z}$, are   effective velocities perpendicular and parallel with respect to z, respectively, and $\tilde k_\pm=(k_x\pm ik_y)/k_0$ with $k_{0}$ being system-dependent constant. Of key importance in this work are the parameters $Q$ and $Q_0$, responsible for breaking time reversal (TRS) and inversion  (IS) symmetries, respectively. The exponent $J$ in Eq.~\eqref{hamiltonian:model} represents the winding number associated with the multi-Weyl/Dirac topological charge~\cite{PhysRevLett.108.266802}. Finally, $\sigma$ and $\tau$ are Pauli matrices acting on the  spin and chirality sectors of the Hilbert space, respectively, with $\sigma_0$ and $\tau_0$ being the corresponding  identities. To simplify the notation, we have introduced $\sigma_{\pm}=\big(\sigma_x \pm i\sigma_y\big)/2$. 

The impurity Hamiltonian in turn can be written as 
\begin{equation}
H_{\rm imp}=\sum_{s=\uparrow,\downarrow}(\varepsilon_{d} -\mu \big) d^{\dagger}_{s}d_{s}+Un_{d\uparrow}n_{d\downarrow},
\label{impurityhamiltonian}
\end{equation}
where $d_{s}^{\dagger} (d_{s})$ creates (annihilates) an electron in the impurity with spin $s$ and energy $\varepsilon_d$ at the impurity site. $n_{d\sigma}=d_{\sigma}^{\dagger}d_{\sigma}$ is the impurity number operator and $U$ is the Coulomb interaction repulsion energy. 

Assuming, for simplicity, that the impurity hybridizes equally with all bands with a $\veck$-independent matrix element $V$~\cite{Vk_note}, the Hamiltonian ${H}_{\rm V}$ can be written as $ H_{\rm V}=\sum_{\mathbf{k}} (\Psi^\dagger_{\mathbf{k}} { \hat V} \Psi_d +{\rm H.c})$, where $\Psi^\dagger_d=(d^\dagger_\up,d^\dagger_\dn)$ and ${ \hat  V}$ is a matrix given by~\cite{PhysRevB.99.115109}
\begin{equation}
{ \hat  V}=
\begin{pmatrix}
V & 0 & V & 0 \\
0 & V & 0 & V
\end{pmatrix}.
\label{hybmatrix}
\end{equation}
\begin{figure*}[!t]
\includegraphics[scale=0.6]{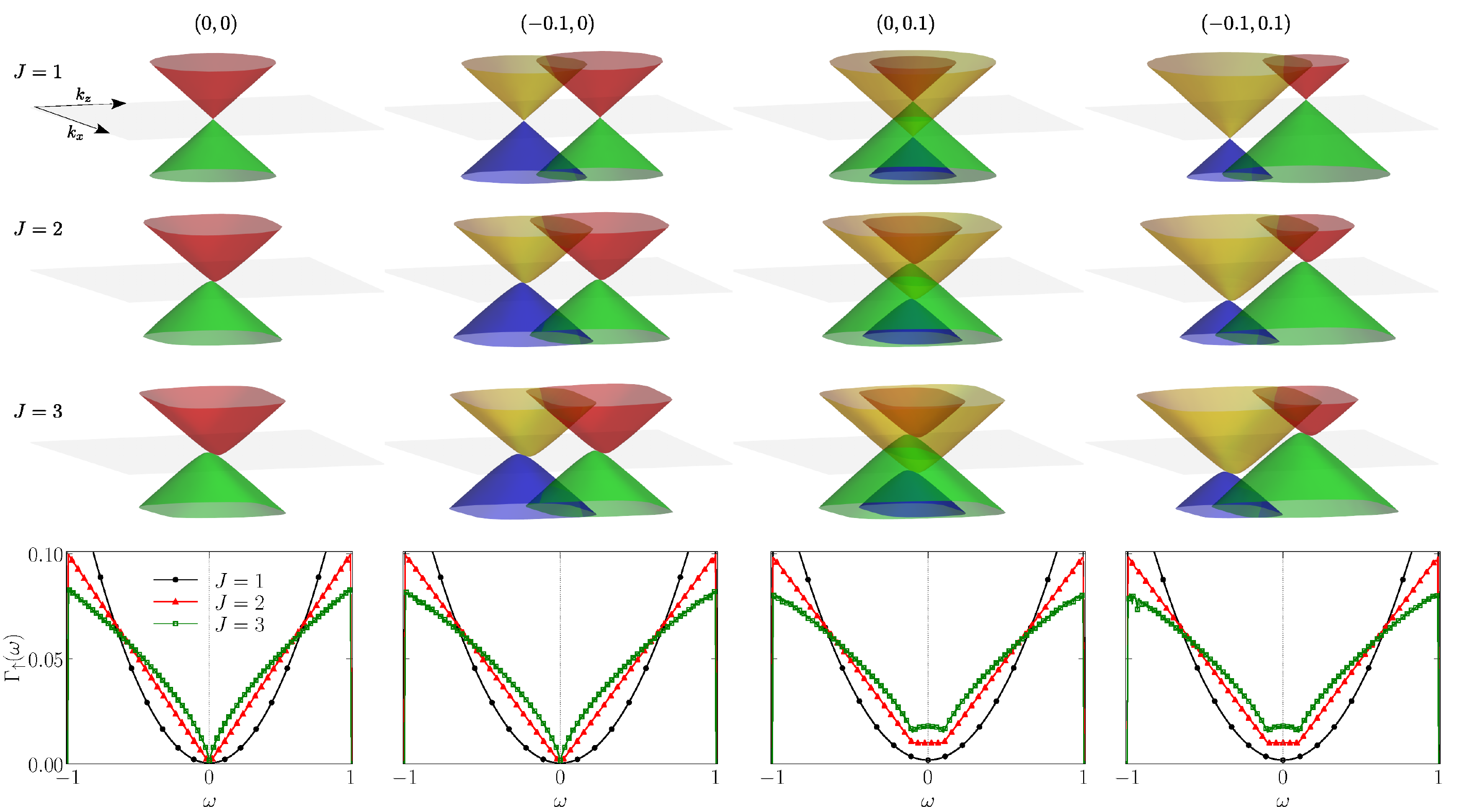}
\caption{Energy bands vs $k_x$ and $k_z$ ($k_z=0$) (top three layers) and  hybridization vs energy for different values of $J$ (lowest layer). The different columns correspond to different values of $Q$ and $Q_0$ indicated by the pairs $(Q,Q_0)$ at the top of the columns. The dashed gray vertical lines in the lowest panels represent the chemical potential, here set at $\mu=0$. Other parameters are $k_0=v_\perp=v_z=1$.   
\label{fig1}
}
\end{figure*}

To access the low-energy physics of the full system we employ the NRG method which allows us to compute the impurity spectral and thermodynamic properties~\cite{Bulla}. Within this method, the entire effect of the host material on the impurity is given by the so-called hybridization function, $\Gamma(\omega)$. To obtain this quantity, we define local Green's function 
\begin{eqnarray}\label{G_imp}
 \hat{ G}_{\rm imp}(\omega) = \left[(\omega -\e_d)\sigma_0 - \hat \Sigma^{(I)}(\omega)-  {\hat \Sigma}^{(0)}(\omega) \right]^{-1},
\end{eqnarray}
where $\hat \Sigma^{(I)}(\omega)$ and $\hat \Sigma^{(0)}(\omega)$ are the interacting and hybridization self-energies. The latter accounts for the effect of the host fermions on the impurity and is formally given by  
\begin{eqnarray}
{\hat \Sigma}^{(0)}(\omega)=\int\dfrac{d\textbf{k}}{(2\pi)^{3}}{\hat V}{G}_{\textbf{k}}^{\rm host}(\omega){\hat V}^{\dagger},
\label{Sigma}
\end{eqnarray}
in which 
\begin{equation}
{\hat G}_{\textbf{k}}^{\rm host}(\omega) = \left[ \omega\mathbb{I} - {\cal H}_{\textbf{k}} \right]^{-1}.
\label{G_host}
\end{equation}
In spherical coordinates, $k_x=k\sin\theta\cos\phi$, $k_y=k\sin\theta\sin\phi$, and $k_z=k\cos\theta$, the Green's function \eqref{G_host} can be recast in the matrix form

\begin{widetext}

\begin{eqnarray}
\hat G^{\rm host}_{\bf k}(\omega)=\left(
\begin{array}{cccc}
 g_{+,Q_0,Q}(k,\theta) & -e^{-i J \phi } F_{Q_0,Q}(k,\theta) & 0 & 0 \\
 -e^{i J \phi }F_{Q_0,Q}(k,\theta) & g_{-,Q_0,Q}(k,\theta) & 0 & 0 \\
 0 & 0 & g_{-,-Q_0,-Q}(k,\theta) & e^{-i J \phi } F_{-Q_0,-Q}(k,\theta) \\
 0 & 0 & e^{i J \phi } F_{-Q_0,-Q}(k,\theta) & g_{+,-Q_0,-Q}(k,\theta) \\
\end{array}
\right).
\end{eqnarray}
In the above we have defined
\begin{eqnarray}
g_{\pm,Q_0,Q}(k,\theta)=\frac{1}{\omega+\mu +Q_0\mp v_z (k \cos \theta -Q)-\frac{k_0^2 v_{\perp}^2 \left(\tilde{k} \sin \theta \right)^{2 J}}{\omega+\mu +Q_0\pm v_z (k \cos \theta -Q) } }
\end{eqnarray}
and
\begin{eqnarray}
F_{Q_0,Q}(k,\theta)=\frac{k_0 v_{\perp} \left(\tilde{k} \sin \theta \right)^J}{k_0^2 v_{\perp}^2 \left(\tilde{k} \sin \theta \right)^{2 J}+v_z^2 (Q-k \cos \theta )^2-\left(\mu +Q_0+\omega \right){}^2}
\end{eqnarray}
\end{widetext}
With these expression, the integrand of Eq.~\eqref{Sigma} becomes
\begin{eqnarray}
{\hat V}\hat{G}_{\textbf{k}}^{\rm host}(\omega){\hat V}^{\dagger}=
{V}^2\begin{pmatrix}
A_{Q_0,Q}(k,\theta)& e^{-iJ\phi}B_{Q_0,Q}(k,\theta)\\
e^{iJ\phi} B_{Q_0,Q}(k,\theta) & A_{-Q_0,-Q}(k,\theta)
\end{pmatrix}\nonumber\\
\end{eqnarray}
where $A_{Q_0,Q}(k,\theta)=g_{+,Q_0,Q}(k,\theta)+g_{-,-Q_0,-Q}(k,\theta)$ and $B_{Q_0,Q}(k,\theta)=F_{-Q_0,-Q}(k,\theta)-F_{Q_0,Q}(k,\theta)$.
Transforming the integral \eqref{Sigma} into spherical coordinates, it is easy to see that the off diagonal terms vanish under integration over the azimuthal angle $\phi$, while the remaining diagonal terms can be written as 
\begin{eqnarray}
{\Sigma}^{(0)}_{s}(\omega)\!\!=\!\!\frac{V^2}{4\pi^2}\int_0^\infty\!\! k^2 dk\int_0^\pi\!\! d\theta \sin \theta A_{sQ_0,sQ}(k,\theta,\omega).
\label{Sigma1}
\end{eqnarray} 
On the rhs of the expression above, $s=+1$ and $-1$ for spin $\up$ and $\down$, respectively. For practical purpose we introduce a cutoff $k_c$ to truncate the integral over $k$ \cite{kc_note}.  

Within the NRG approach, the effect of the host material on the impurity and entirely  accounted for by \textit{hybridization function}  defined as
\begin{eqnarray}
\Gamma_s(\omega)= -{\rm Im}[\Sigma^{(0)}_{s}(\omega+i 0^+)].
\label{Deltafunction}
\end{eqnarray}
The integration in Eq.~\eqref{Sigma1} is rather complicated and, in general, cannot be performed analytically. Therefore, we solve it numerically. Having the numerical results for $\Gamma_s(\omega)$ we apply the standard numerical logarithmic discretization scheme described in Ref.~\cite{Bulla}.

To get some insight on what we expect from our numerical results for the Kondo physics, it is instructive to take a close look at the structure of the hybridization function in the various regimes of the Hamiltonian model. This will allow us to compare the Kondo regimes expected here with those already known in the literature.  
For a given set of parameters of the host Hamiltonian, the contribution for $\Gamma_s(\omega)$ is associated to the four energy bands of the clean multi-Weyl/Dirac node system [obtained upon diagonalizing the Hamiltonian \eqref{hamiltonian:model}],
\begin{eqnarray}
\e_{hs\veck}=h\sqrt{k_0^2 v_{\perp}^2 {\tilde k_{\perp}}^{2 J}+v_z^2 \left(Q-sk_z\right){}^2} -sQ_0-\mu.
\end{eqnarray}
Here, $k_0^2\tilde k_\perp^2={k_x^2 + k_y^2}$, $h$ represents conduction (+) and valence (-) bands and  $s=\pm$ is the quantum number resulting from linear combinations of the original spin and chirality quantum numbers.  These bands are shown in the three upper layers of Fig.~\ref{fig1}. Each of these layers corresponds to a given value of $J$ indicated on the left hand side of the figure, while a given set of $Q$ and $Q_0$ is indicated by the pair $(Q,Q_0)$ placed at the top of each column. The lower layer of Fig.~\ref{fig1} shows the hybridization function for different representative instances of the multi-Weyl and Dirac node cases studied here. In the next section we will discuss in detail the various features of the hybridization function alongside the discussion of the Kondo physics dredged up by the NRG calculations.

\section{Numerical results} \label{numerical_results}

To perform our numerical analysis, we will look mainly at the impurity magnetic moment $k_B T\chi_{\rm imp}(T)$ and the local density of states $\rho(\omega)$ that reveal most of the relevant features of the Kondo effect. Here, $\chi_{\rm imp}(T)$ is the impurity contribution to the magnetic susceptibility defined as $\chi_{\rm imp}(T)\equiv\chi(T)-\chi^{(0)}(T)$, where $\chi(T)-\chi^{(0)}(T)$ are the magnetic susceptibilities calculated with and without the impurity, respectively (see discussion in Ref.~\cite{Bulla}), and $\rho_s(\omega)$ can be defined using the diagonal elements of the impurity local Green's function matrix \eqref{G_imp} as $\rho_s(\omega)=-\frac{1}{\pi}{\rm Im}[\hat{G}_{{\rm imp}}(\omega + i0^{+})]_{ss}$. Within the NRG approach, $\rho(\omega)$ is calculated via Lehman representation $\hat{G}_{{\rm imp}}(\omega + i0^{+})]_{ss}$ using the many-body spectrum readily available in standard  NRG calculations. For more details, see thorough discussion in Ref.~\cite{PhysRevB.79.085106}.
To obtain the numerical results, we set $D=1$ as the ultraviolet energy cutoff for all the calculations which follows. Moreover, we set the model parameters $v_\perp=v_z=k_0=1$. For simplicity, we also set the fundamental Boltzmann and Plank constants $k_B=\hbar=1$. We will fix the impurity-related parameters $V=0.12625$, $U=0.2$ and $\varepsilon_d=-U/2$. There is nothing special with the specific choice of $V$ here. It just renders $\Gamma_\up(D)=\Gamma_\dn(D)=0.1$, which in combination with $U=0.2$ and $\epsilon_d=-U/2$ (corresponding to the particle-hole symmetric point of the impurity Hamiltonian) provides us with $T_K$ much smaller than $D$, $U$, and $\Gamma_s$ for any regime considered in this work. This prevents the system from entering in a mixed-valence regime.
Our NRG calculations were obtained using the well-known  open source NRG Ljubljana code~\cite{PhysRevB.79.085106}, using $\Lambda=2.5$, keeping a maximum number of states $N_{s}=2000$ at each iteration and z-averaging $N_z=4$. The NRG flow is truncated at a lowest energy $E_{\rm min}=10^{-12}$. This choice is good enough to access the main feature of the Kondo regime with great accuracy. Finally, the impurity density of states $\rho_s$ was calculated at $T=10^{-10}$.


\subsection{Multi-Dirac node fermions: $Q=Q_0=0$}

Let us start by studying the regime of $(Q,Q_0)=(0,0)$, in which the Hamiltonian \eqref{hamiltonian:model} is invariant under TRs and IS. The Hamiltonian then describes the aforementioned  multi-Dirac node materials \cite{Yang2014,PhysRevLett.115.036807,WU2019125876}. For $\mu=0$ and $\mu\neq 0$ it corresponds to semimetal and metal, respectively. 
The energy bands and hybridization functions for this case are shown in the first column of Fig.~\ref{fig1}. For $J=1$ the host corresponds to a four-band Dirac  semimetal (for $\mu=0$) exhibiting two copies of Dirac cones. This renders a hybridization function that behaves as  $\Gamma(\omega)\equiv\Gamma_\up(\omega)\sim |\omega-\mu|^2 = \Gamma_\dn(\omega)$ (since  TRS is preserved). For $J=2,3$, the bands are distorted around $\veck=0$. Note that, while $\e(0,0,k_z)\sim |k_z|$ the dispersion is no longer linear with $k_x$ or $k_y$. The hybridization function behaves as $\Gamma(\omega)\sim |\omega-\mu|$ and $\Gamma(\omega) \sim |\omega-\mu|^{2/3}$ (for $J=2$ and $J=3$, respectively). These results are consistent with the prediction that $\Gamma(\omega) \sim |\omega-\mu|^{2/J}$ as $\omega \rightarrow 0$ for a generic $J$ ~\cite{PhysRevB.93.201302, PhysRevB.95.201102}.  

\begin{figure}[!t]
\begin{tabular}{c}
\includegraphics[scale=0.35]{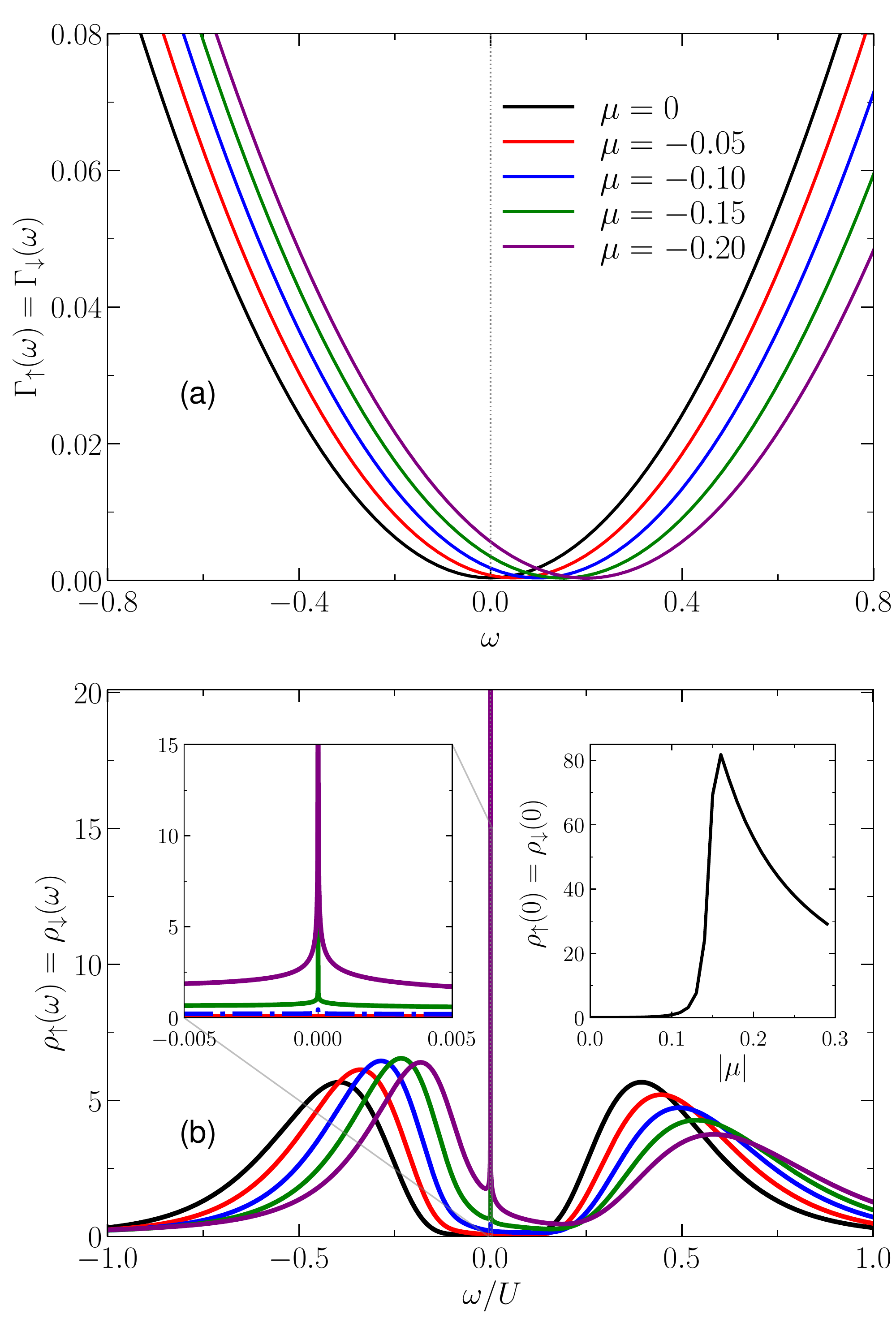}
\end{tabular}
\caption{Hybridization function (a) and impurity density of states $\rho_{s}$ ($s=\up , \dn$) (b) as  a function of $\omega$ for $J=1$ and for different values of $\mu$. The left inset of (b) shows a zoom of the region about $\omega=0$, while the right inset shows the height of the peak, $\rho_{s}(0)$, as a function of $|\mu|$.}
\label{fig2}
\end{figure}

The Kondo physics emerging from this class of hybridization functions is well understood as it has been thoroughly investigated previously in a generic context, namely the so-called pseudogap regime~  \cite{PhysRevLett.64.1835}. Indeed, the Anderson and Kondo models for a 
pseudogap density of states $\rho(\omega) \propto |\omega|^{r}$ presents a rich quantum phase diagram extensively studied~ \cite{PhysRevLett.89.076403,PhysRevB.70.214427,PhysRevB.74.144410,PhysRevB.96.045103,Bulla1997,PhysRevB.88.195119,PhysRevB.72.045117} [remember that $\Gamma(\omega)\propto \rho_{\rm host}(\omega)$]. Numerical and perturbative renormalization group calculations showed that the fixed points structure of the pseudogap Anderson/Kondo systems are radically distinct for $r<1$ and $r>1$, suggesting $r=1$ as the upper-critical “dimension” of the problem in the RG sense \cite{PhysRevLett.89.076403,PhysRevB.70.214427}. For completeness, here we present the results we have obtained for these three regimes when $J=1,2$, or $3$. From a theoretical point of view, cases of $J>3$ are interesting as they render $r\leq 1/2$, leading to anomalous Kondo screenings~\cite{PhysRevB.57.14254}.  It has been shown, however, that Weyl materials with $J> 3$ are not topologically protected~\cite{PhysRevLett.108.266802}. Therefore, these cases are not considered here. The results for $J=1$, which corresponds to 
$r=2$, is summarized in Fig.~\ref{fig2}. 
In Ref.~\cite{Vojta_2010}, the authors have  showed that the Kondo screening in this case is influenced by the electron-hole asymmetry of the pseudogap density of states. Here, the electron-hole asymmetry of the multi-Dirac/Weyl hosts is controlled by the chemical potential $\mu$, resulting in significant changes in the low-energy physics in comparison with the electron-hole symmetric case~($\mu=0$). This is shown in Fig.~\ref{fig2}(a) that shows $\Gamma(\omega)$ vs $\omega$ for various values of $\mu$. Note that the hybridization function vanishes quadratically at $\omega=-\mu$, which results in a finite hybridization function at $\omega=0$. This is why for $\mu=0$ there is no Kondo peak, as seen in Fig.~\ref{fig2}(b) (black line). Note also that as $\Gamma(0)$ increases with  $|\mu|$ the system becomes metallic,  resulting in the emergence of the Kondo peak (note the sharp peaks for $\mu=-0.15$ and $\mu=-0.2$). The finite Kondo temperature ($T_K$) in this case is known in the literature \cite{HewsonBook,Mitchell}. The right inset of Fig.~\ref{fig2}(b) shows the evolution of $\rho_{s}(0)$ (the height of the impurity density of states) with $|\mu|$. Observe that the Kondo peak starts increasing very rapidly for $|\mu| \approx 0.1$. For very small $|\mu|$, although finite,  $T_K$ is smaller than $T$ (the temperature at which $\rho_s$ is calculated. The maximum value of $\rho_{s}(0)$ for $|\mu|\approx 0.16$ suggests the complete onset of the Kondo screening when $T_K$ becomes larger than $T$, after which  the broadening of the Kondo resonance is more pronounced.

\begin{figure}
\begin{tabular}{c}
\hskip-0.5cm
\includegraphics[scale=0.35]{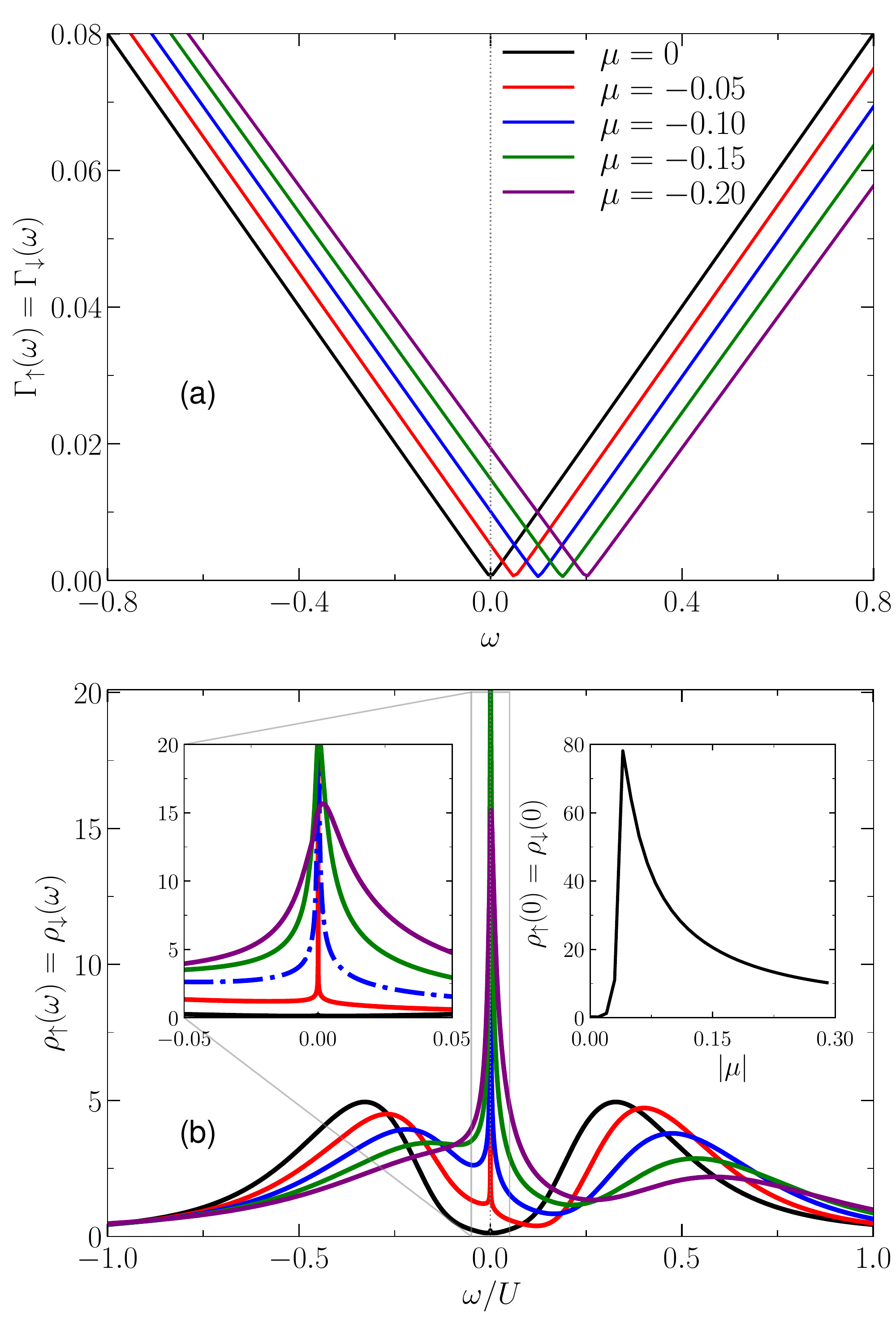}
\end{tabular}
\caption{
Hybridization function (a) and impurity density of states $\rho_{s}$ ($s=\up , \dn$) (b) as  a function of $\omega$ for $J=2$ and for different values of $\mu$. The left inset of (b) shows a zoom of the region about $\omega=0$, while the right inset shows the height of the peak $\rho_{s}(0)$, as a function of $|\mu|$.
}
\label{fig3}
\end{figure}

\begin{figure}
\begin{tabular}{c}
\includegraphics[scale=0.35]{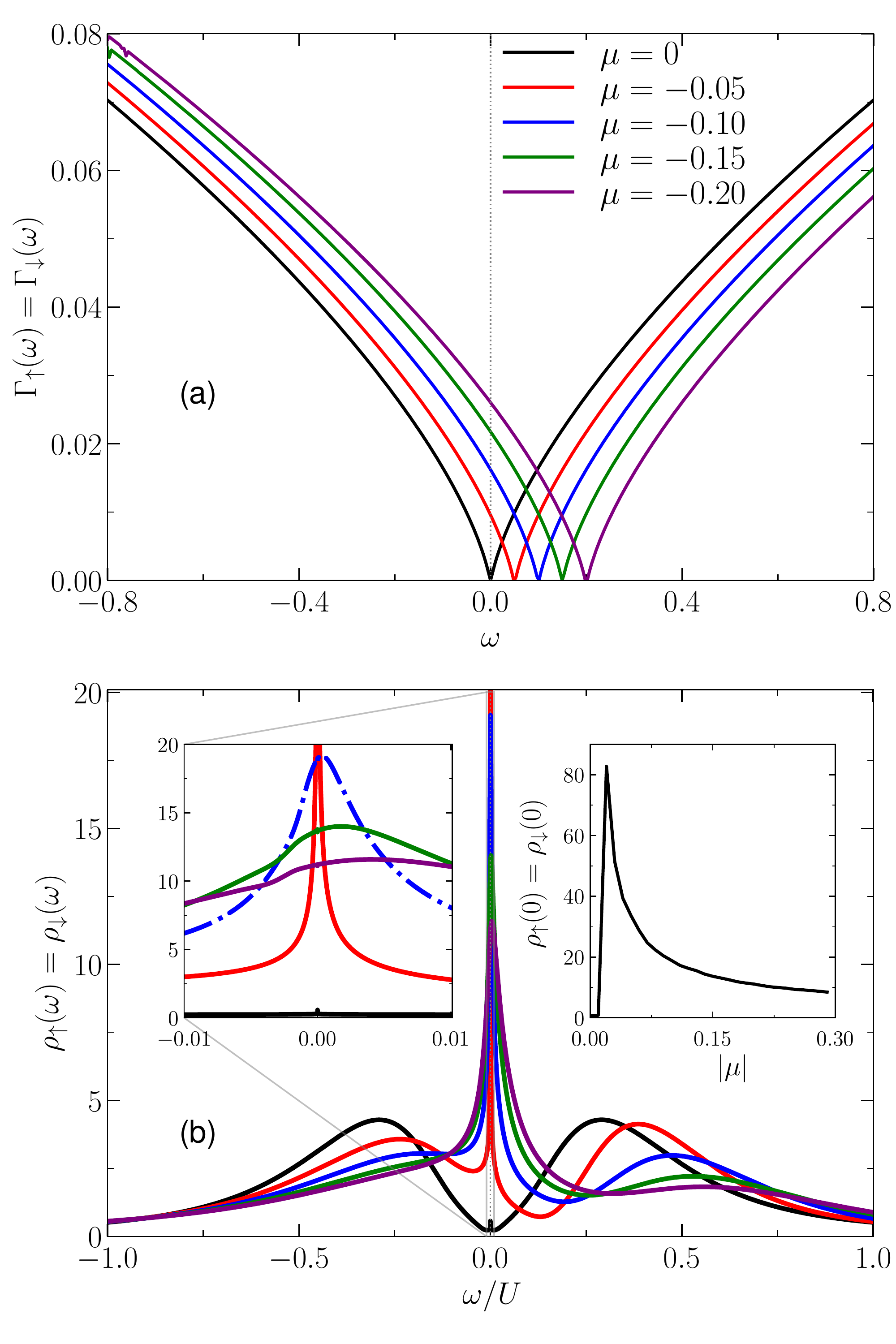}
\end{tabular}
\caption{
Hybridization function (a) and impurity density of states (b) as  a function of $\omega$ for $J=3$ and for different values of $\mu$. Left inset of (b) shows a zoom of the region about $\omega=0$, while right inset shows the height of the peak, $\rho_{s}(0)$, as a function of $|\mu|$.
}
\label{fig4}
\end{figure}
The result for $J=2$ and $J=3$ ($r=1$ and $r=2/3$) are shown in Figs.~\ref{fig3} and \ref{fig4}, respectively, showing the same quantities of Fig.~\ref{fig2}. Note that the results for the three values of $J$ are qualitatively equivalent. This equivalence is consistent with the predictions of Ref.~\cite{PhysRevB.96.045103} that identify all cases of $r>1/2$ as belonging to same class of pseudogap Kondo screening. Note, however that for larger values of $J$ the Kondo peaks is broader for a fixed value of $\mu$. This can be understood by noticing that for a given value of $\mu$ and any $\omega\neq\mu$, $\rho_{\rm host}(\omega)=|\omega -\mu|^{2/J}$ is larger for larger $J$. Hence, the impurity is more strongly hybridized with the host material for larger $J$. It is also noteworthy that  the value of the chemical potential below which $\rho_{s}(0)$ drops to zero is smaller for larger $J$. We will come back to this point below.

\begin{figure}
\begin{tabular}{c}
\includegraphics[scale=0.53]{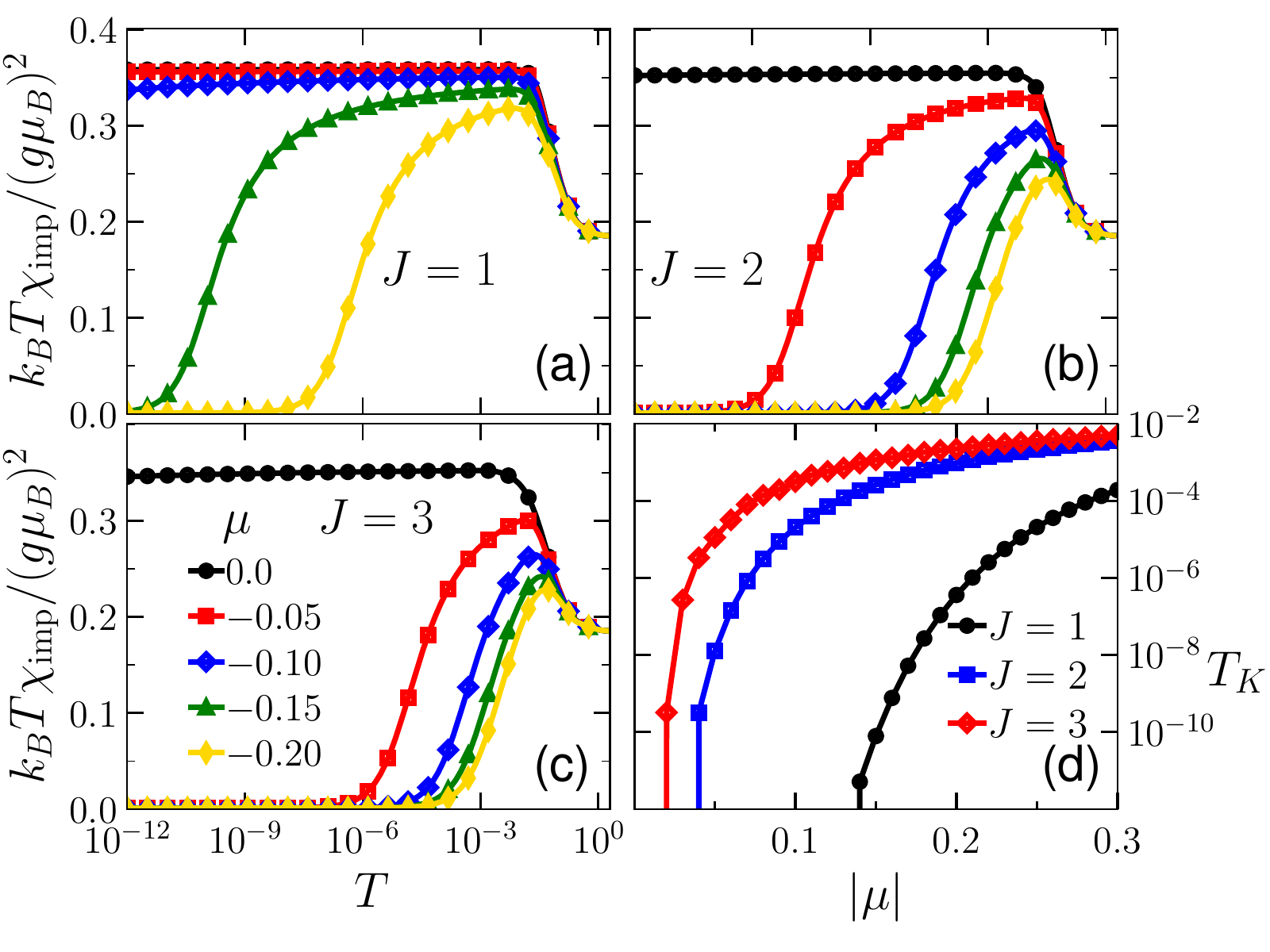}
\end{tabular}
\caption{Impurity magnetic moment for $J=1$ (a), $J=2$ (b), and $J=3$ (c) and various values of $\mu$. (d) Kondo temperature as a function of $\mu$ for different values of $J$. Other parameters are $Q=Q_0=0$.}
\label{fig5}
\end{figure}

To conclude this section, in Fig.~\ref{fig5}(a), \ref{fig5}(b), and \ref{fig5}(c) we show the impurity magnetic moment, $k_BT\chi_{s}$ as a function of temperature for $J=1$, $J=2$, and $J=3$, respectively, and various values of $\mu$. Note that for all $J$ and $\mu=0$ the magnetic moment remains finite all the way to $T\rightarrow 0$, confirming the doublet ground state, a characteristic of the LM fixed point. However, for $\mu\neq 0$, it drops to zero when $T\rightarrow 0$, which is consistent with the singlet Kondo ground state, the strong coupled fixed point. Figure \ref{fig5}(d) shows $T_K$ as a function of $\mu$ for the three values of $J$~\cite{Tk_note}. Note that $T_K$ becomes vanishingly small as $|\mu| \rightarrow 0$ for all three values of $J$. When $|\mu|$ increases the curves tend to coincide with each other, which is better observed for $J=2,3$. This is an expected behavior since there are two regions along $\omega$ within which the  hybridization functions, $\Gamma(\omega)$, for different $J$ have similar values, see lower panel in the left column of  Fig.~\ref{fig1}.  Interestingly, the values of $|\mu|$ for which $T_K$ crosses $T$ ($=10^{-10}$) agree fairly well with those at which  $\rho_s(0)$ is maximum in the insets of Figs.~\ref{fig2}(b), \ref{fig3}(b) and \ref{fig4}(b), confirming that, indeed, $\rho_s(0)$ vanishes as $|\mu|\rightarrow 0$ because $T_K$ becomes smaller than $T$.

\subsection{Multi-Weyl node fermions: $Q\neq 0$ or  $Q_0\neq 0$}
Up to now we have considered situations in which the system exhibits both TRS and IS invariance. Let us now focus on the Weyl fermions, obtained when at least one of these symmetries is broken.
\subsubsection{TRS-broken multi-Weyl node semimetal  ($Q\neq 0$)}  
\begin{figure}[!t]
\includegraphics[scale=0.504]{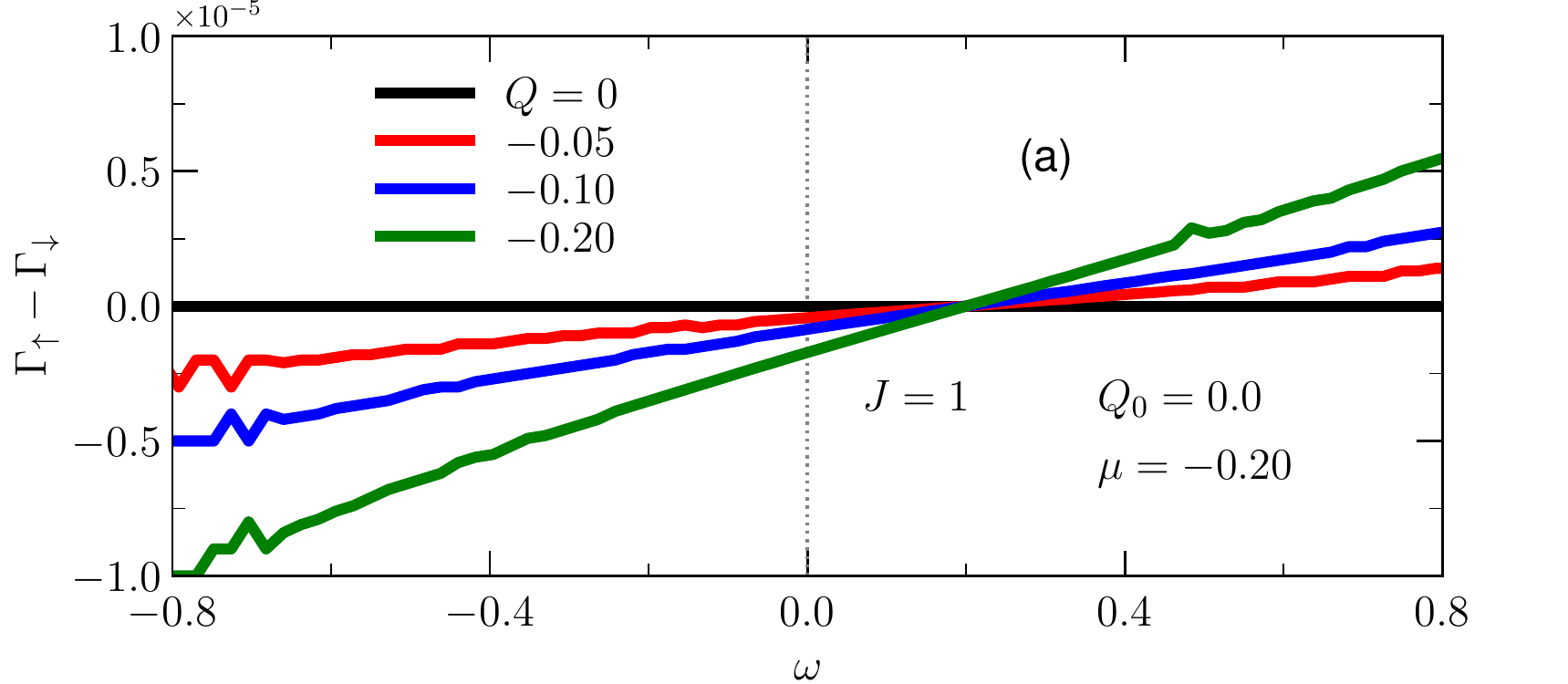}
\includegraphics[scale=0.55]{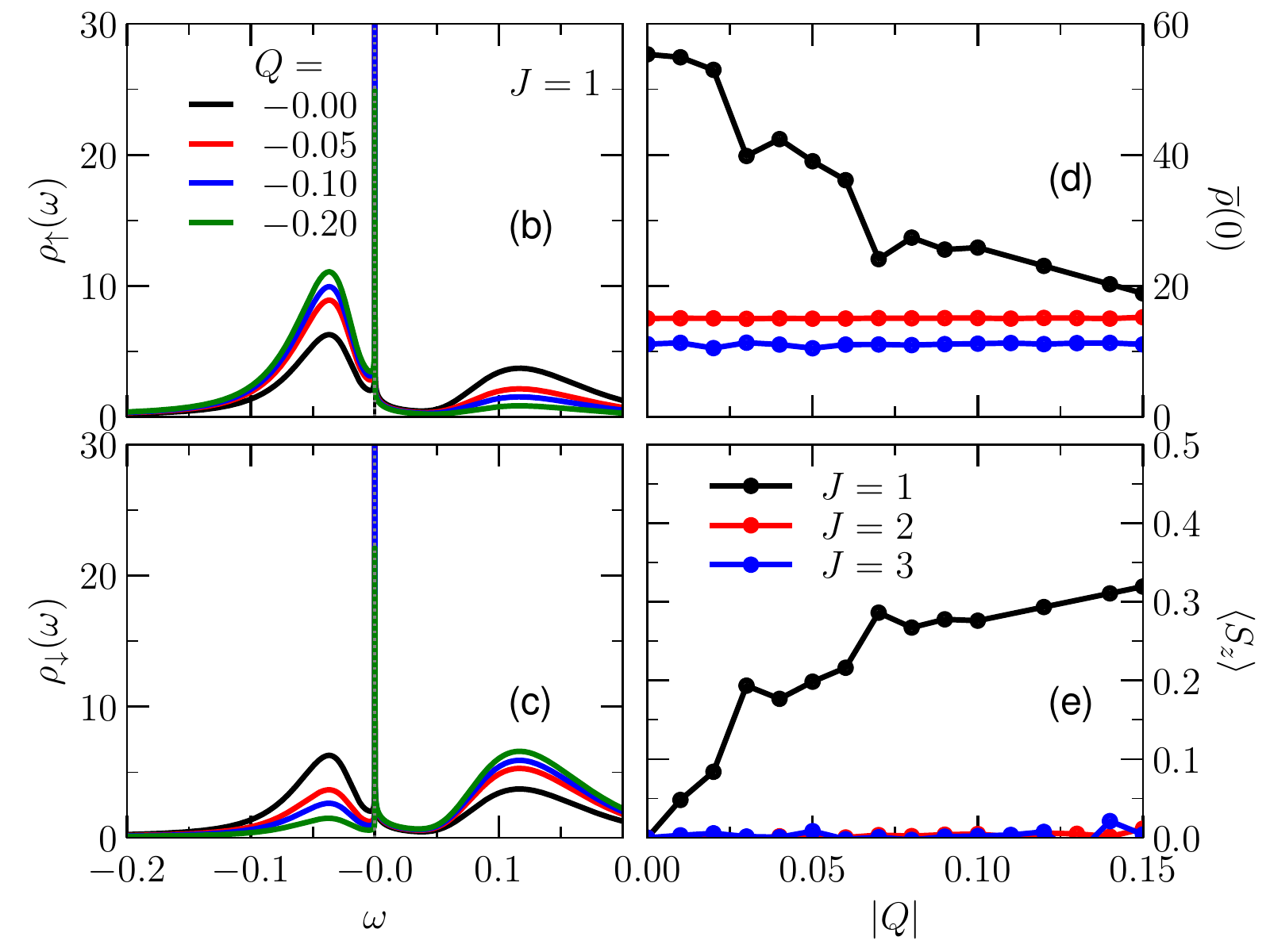}
\caption{(a) Splitting of the hybridization function for several values of $Q$ for a fixed chemical potential $\mu=-0.2$.
Impurity density of states for spin \emph{up} (b) and \emph{down} (c) for the same values of $Q$ as in panel (a). (d)  height of the Kondo peak averaged over spins [$\bar\rho(0)=[\rho_\up(0)+\rho_\dn(0)$]/2 vs $Q$ for $\mu=-0.2$. (e) $\langle S_z\rangle$ vs $|Q|$ also for $\mu=-0.2$. In all panels $Q_0=0$ and  J=1.}
\label{fig:mag}
\end{figure}
%
Here we analyze the situation where only TRS is broken by keeping $Q_0 = 0$. The energy bands and hybridization function for this case is shown in the second column of Fig.~\ref{fig1}, where we used the notation $(Q,Q_0)=(-0.1,0)$. 
The effect of breaking TRS in the energy bands is to shift the nodes along $k_z$. We should note that the conduction (and valence) energy bands have opposite spin polarization. As a result, since here we have assumed that the spins of the impurity couples equally to all bands of the host material, at a given value of $\mu$ the spin splitting of the hybridization functions vanishes upon integration over the entire momentum space. In our approach, the introduction of the momentum cutoff $k_c$ induces a small spin splitting in the hybridization function for a finite $Q$. 
The polarization $\Gamma_\up - \Gamma_\dn$ of the hybridization function is shown in Fig.~\ref{fig:mag}(a) for $J=1$,  $\mu=-0.2$, and several values of $Q$. Observe that it depends on $\omega$ and vanishes at $\omega=-\mu$. This is quite different from that obtained in Ref.~\cite{Li2019} in the context of magnetic graphene, in which $\Gamma_\up(\omega)$ and  $\Gamma_\dn(\omega)$ vanish at different energies, rendering a full polarization of the hybridization function. Here, full polarization is absent and is indeed very small. Nevertheless, it is enough to induce a visible spin splitting in the impurity density of states. In real systems, integration over the well defined first Brillouin zone should naturally induce  spin splitting in the hybridization function. This can be confirmed by adopting  a more realistic model for the  multi-Dirac/Weyl node host material. Moreover, spin-dependent coupling of the impurity orbital to the host bands can also contribute to spin polarization of the hybridization function. This important improvement in the model is beyond the scope of the present work and will be presented in a future publication. 
 
The spin resolved LDOS is shown in Fig.~\ref{fig:mag}(b) (for spin $\up$) and Fig.~\ref{fig:mag}(c) (for spin $\down$) for $J=1$.  Interestingly, despite the relatively large spin  splitting of the LDOS, the suppression of the Kondo peak is slow as $Q$ increases as shown in Fig.~\ref{fig:mag}(d). The spin splitting in the LDOS induces a  sizable magnetization $\langle S_z\rangle$, shown in Fig.~\ref{fig:mag}(e). For $J=2$ and $J=3$, the spin polarization is vanishingly small and is not observed in our calculations [see red and blue curves in Fig.~\ref{fig:mag}(e)]. Moreover, the Kondo temperature is larger for these cases, making the Kondo screening more robust against TRS breaking. A change $Q\rightarrow -Q$ in the above leads to opposite polarization in the hybridization function with identical effect in the Kondo screening.

\subsubsection{IS-broken multi-Weyl node semimetal ($Q_0\neq0$)} 
The situation is radically different for finite $Q_{0}$.  The hybridization function, $\Gamma(\omega)$, is always finite even for $\mu=0$, as shown in the lower panel of the third column of Fig.~\ref{fig1}. Consequently, the system can exhibit Kondo screening for any chemical potential $\mu$.
\begin{figure}[!t]
\includegraphics[scale=0.5]{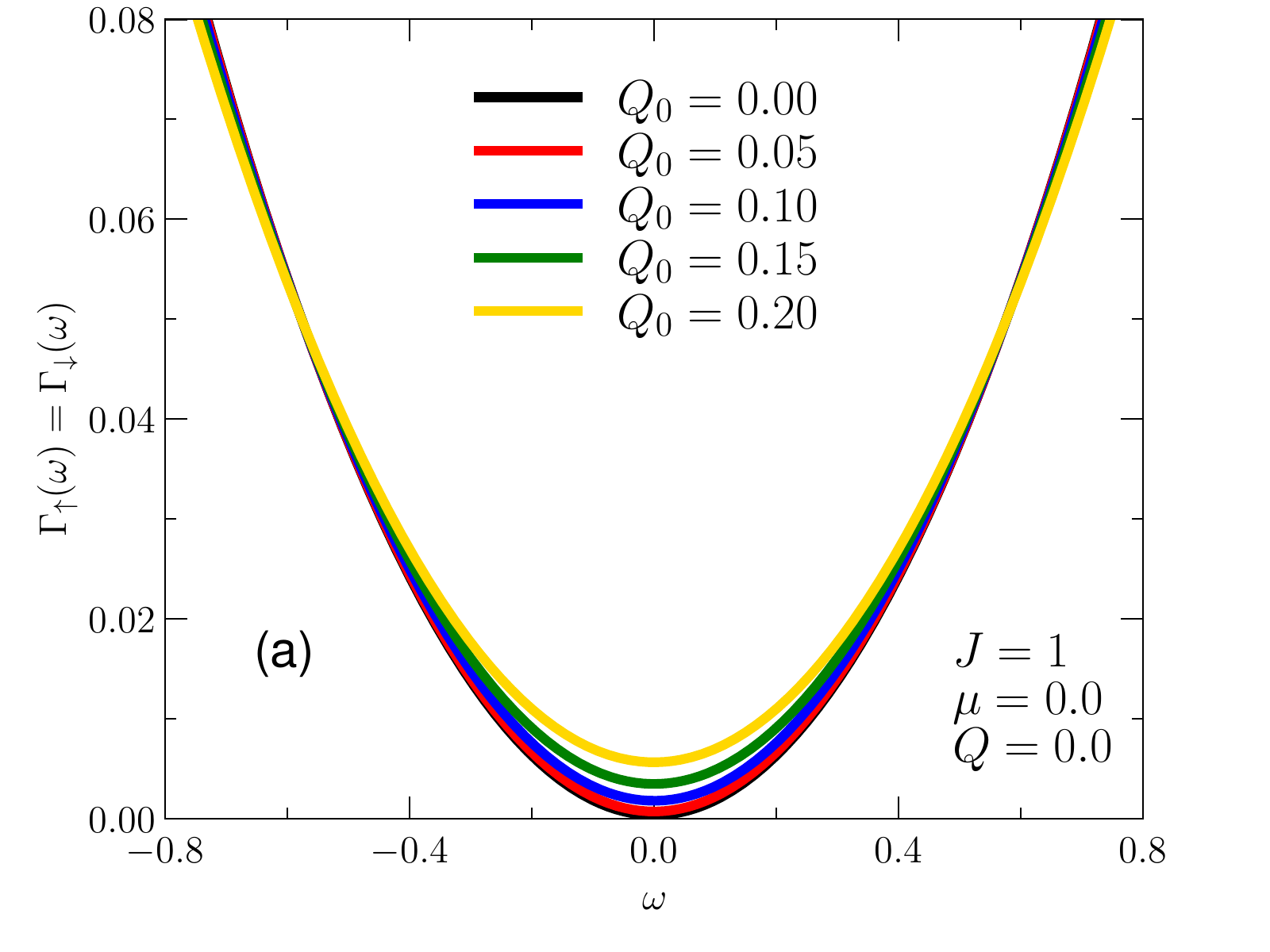}
\includegraphics[scale=0.5]{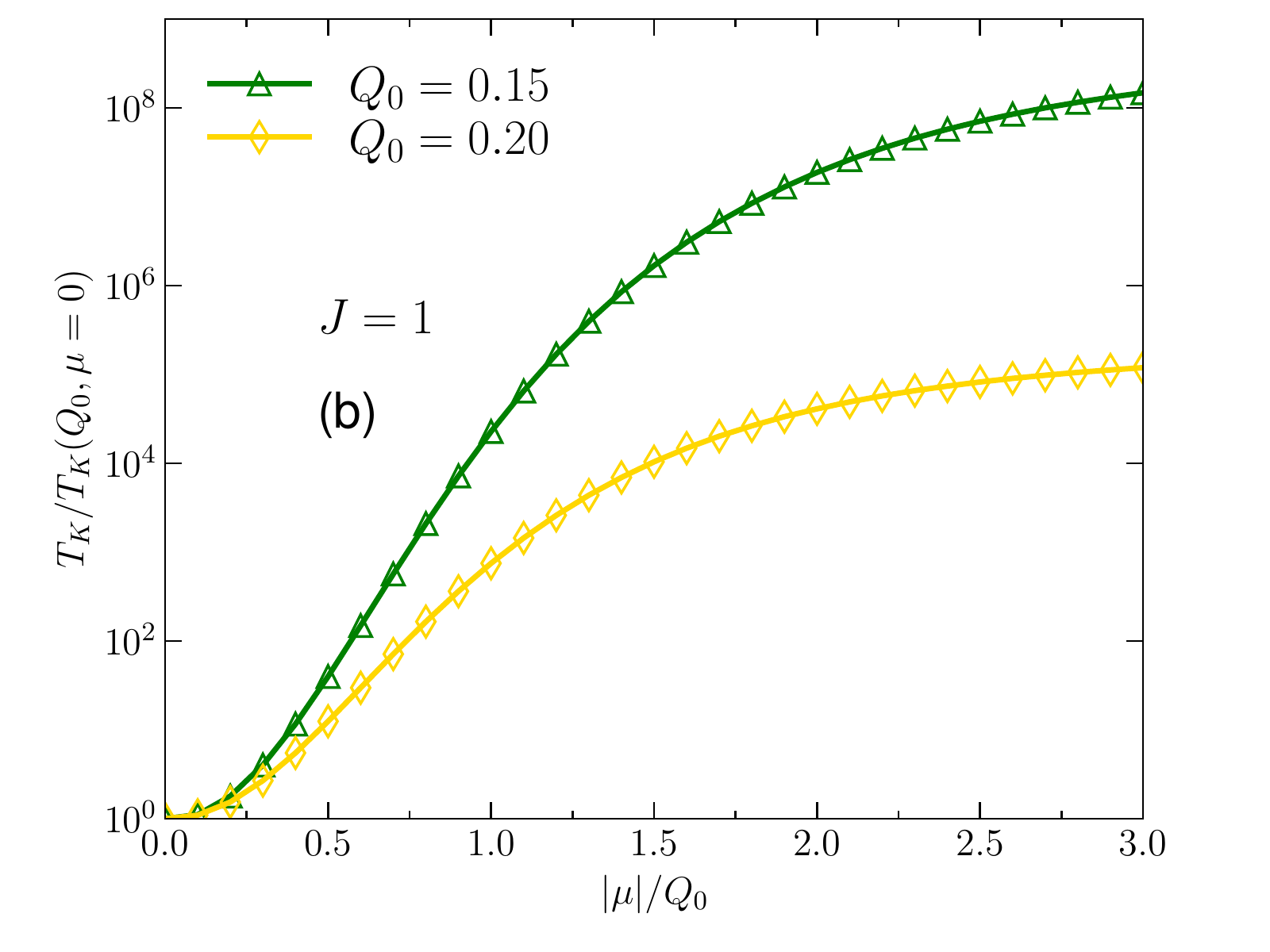}
\includegraphics[scale=0.5]{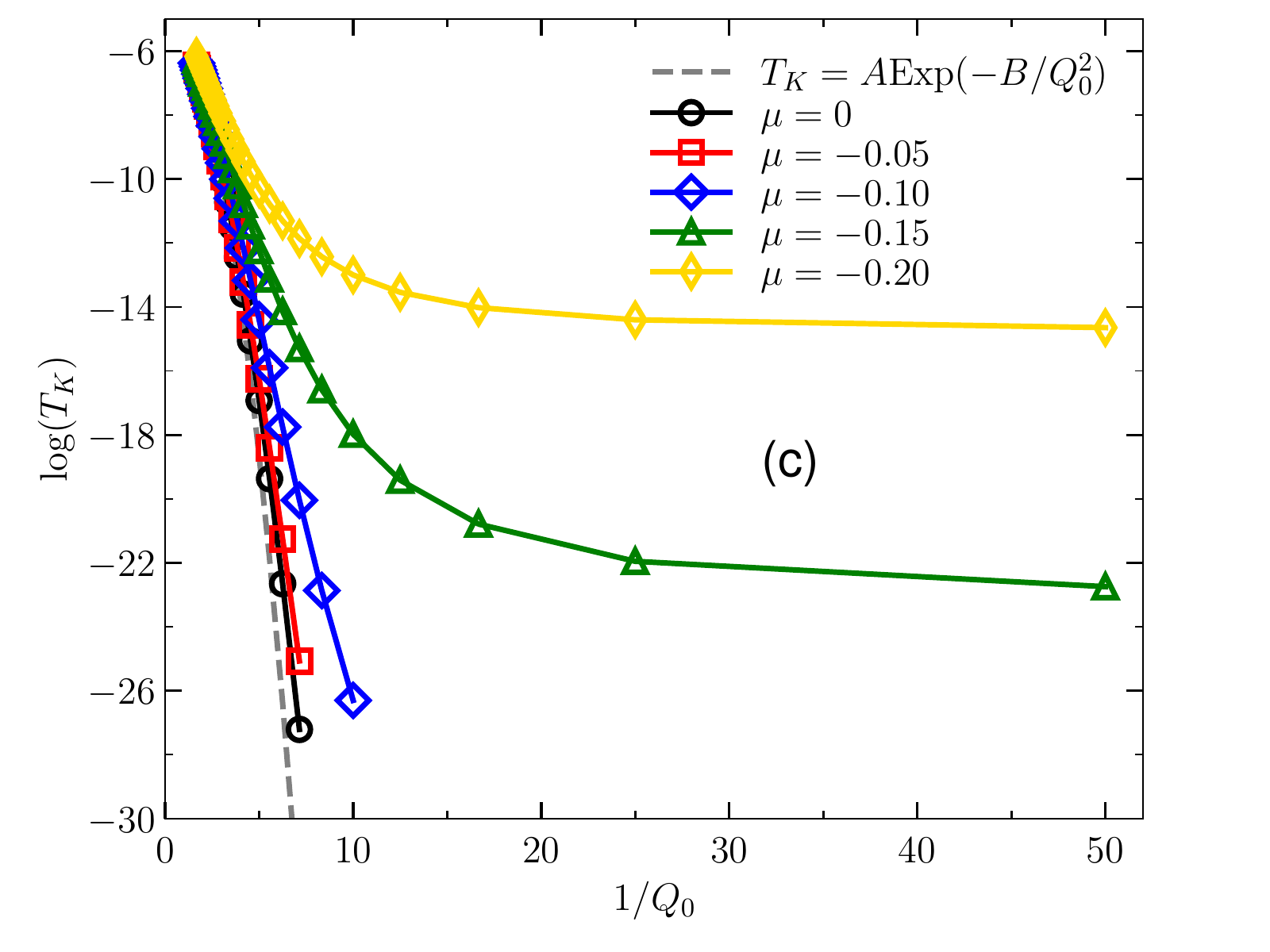}
\caption{(a) Hybridization function vs energy for $J=1$ and various values of $Q_0$. (b) $T_K(Q_0,\mu)/T_K(Q_0,\mu=0)$ as a function of $\mu/Q_0$, for different values of $Q_0$. (c). Kondo temperature vs $Q_0^{-1}$ for various values of $\mu$. Here, $Q=0$ for all panels.}
\label{fig:T_k_mu_J=1}
\end{figure}

Figure \ref{fig:T_k_mu_J=1}(a) shows the hybridization function as a function of $\omega$ for $J=1$, $\mu=0$, and various values of $Q_0$. Note that the parabolas are shifted upwards and becomes slightly flatter as $Q_0$ increases \cite{hybridization_note}.  The progressive enhancement  of the hybridization function at $\omega=0$ as $Q_0$ increases provides the condition for the Kondo screening to take place at finite temperature. In  Fig.~\ref{fig:T_k_mu_J=1}(b) we show the $T_K(Q_0,\mu)/T_K(Q_0,\mu=0)$ as a function of $\mu$ for $Q_0=0.15$ and $Q_0=0.2$, which shows how $T_K$ increases with $|\mu|$. Despite of their values, note that the curves exhibits very similar behavior. This is because a finite value of $\mu$ produces a rigid shift of the curves (not shown) enhancing the value of $\Gamma(0)$ due to more conduction electron's density of states at the Fermi level. The reader may have already anticipated this by noticing that the hybridization function is a smooth function of $\omega$. It is worth mentioning that for $Q_0=0.1$ and $Q_0=0.05$ shown in Fig.~\ref{fig:T_k_mu_J=1}(a), $T_K$ becomes too small to be captured by our NRG flow which was truncated at a lower temperature $T_{\rm min}=10^{-12}$. We examine now, how the Kondo temperature changes with $Q_0$. To see this, in Fig.~\ref{fig:T_k_mu_J=1}(c) we show $T_{K}(Q_0,\mu)$ vs $1/Q_{0}$ for several values of $\mu$. Note that in the limit of large $Q_{0}$ (small $1/Q_0$) all curves tend to collapse onto a single function of $Q_{0}$ given $T_{K}=A{\rm Exp}(-B/Q_{0}^{2})$, where $A$ and $B$ depend on the parameters of the system. This is because in this limit the density of states of the effective conduction band depends essentially on $Q_{0}$ and becomes $\rho_{\rm host}(\omega)\sim Q_{0}^{2}$ even for finite values of $\mu$~\cite{hybridization_note}, and the Kondo temperature approaches to its ``standard" form for usual metallic systems $T_{K}\sim {\rm Exp}[-1/\rho(0)J_0]$~\cite{HewsonBook}, where $J_0$ is the Kondo coupling of the equivalent effective low-energy Kondo model~\cite{KM_note}.

\begin{figure}[!t]
\includegraphics[scale=0.5]{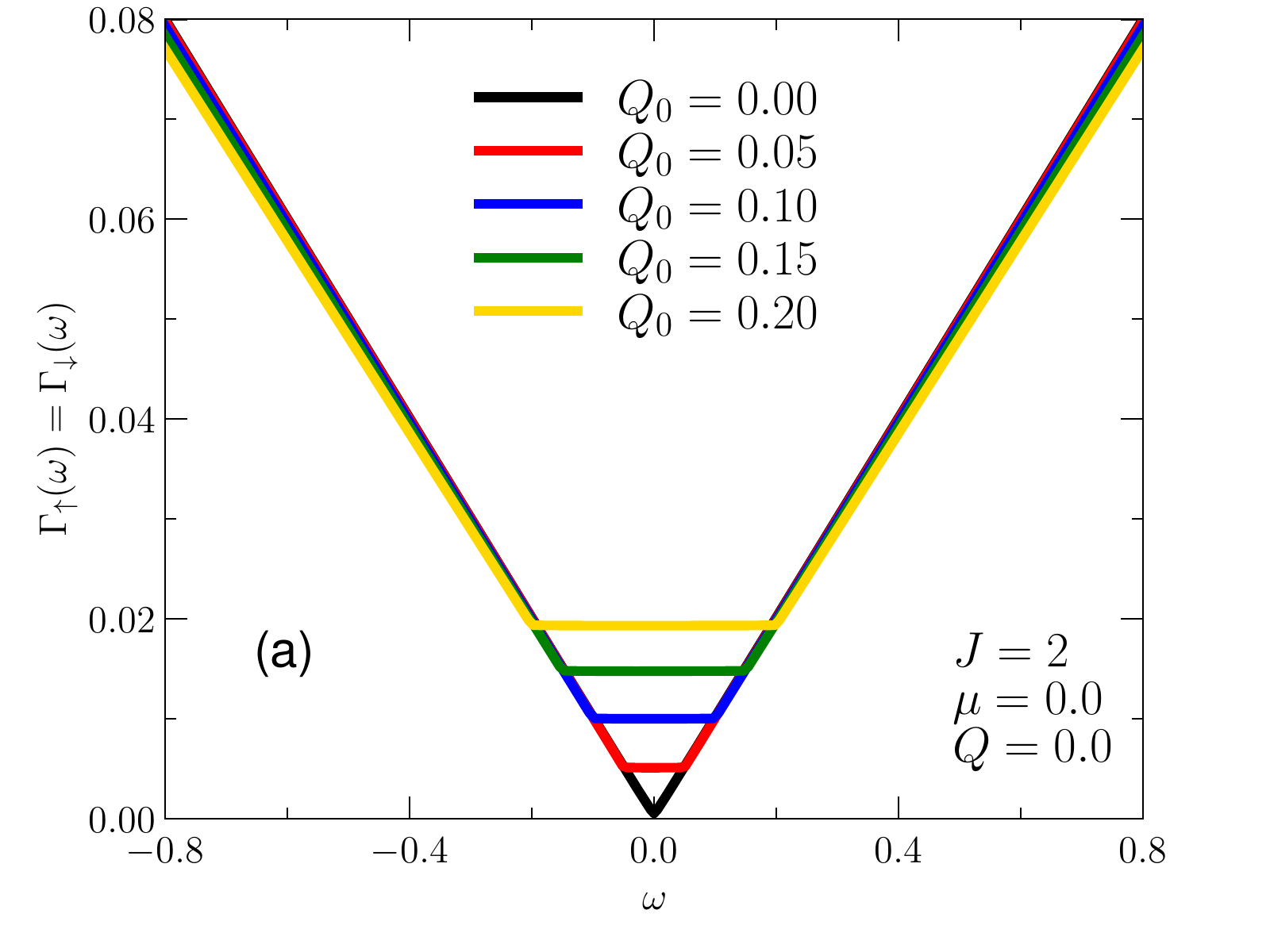}
\includegraphics[scale=0.5]{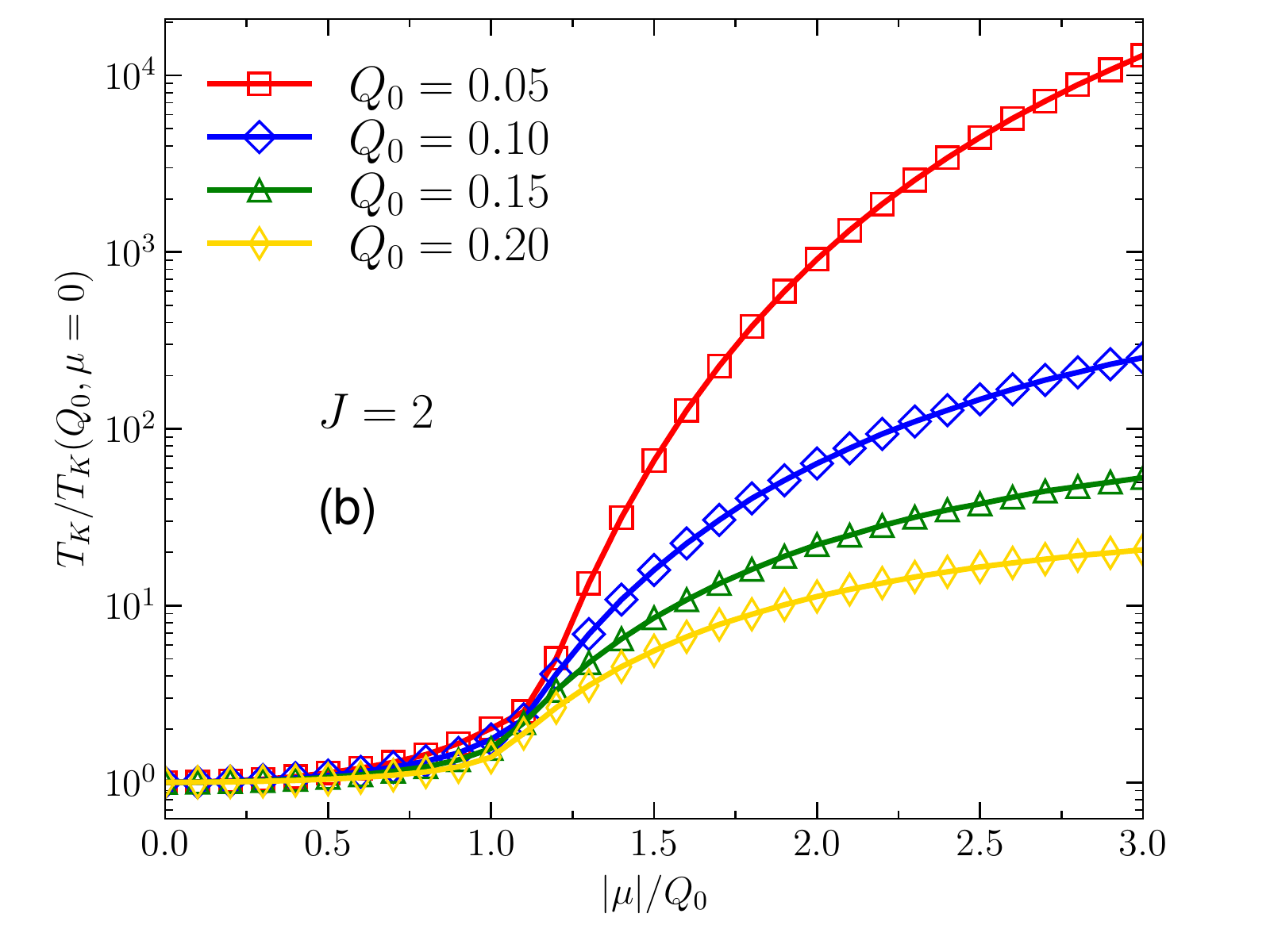}
\includegraphics[scale=0.5]{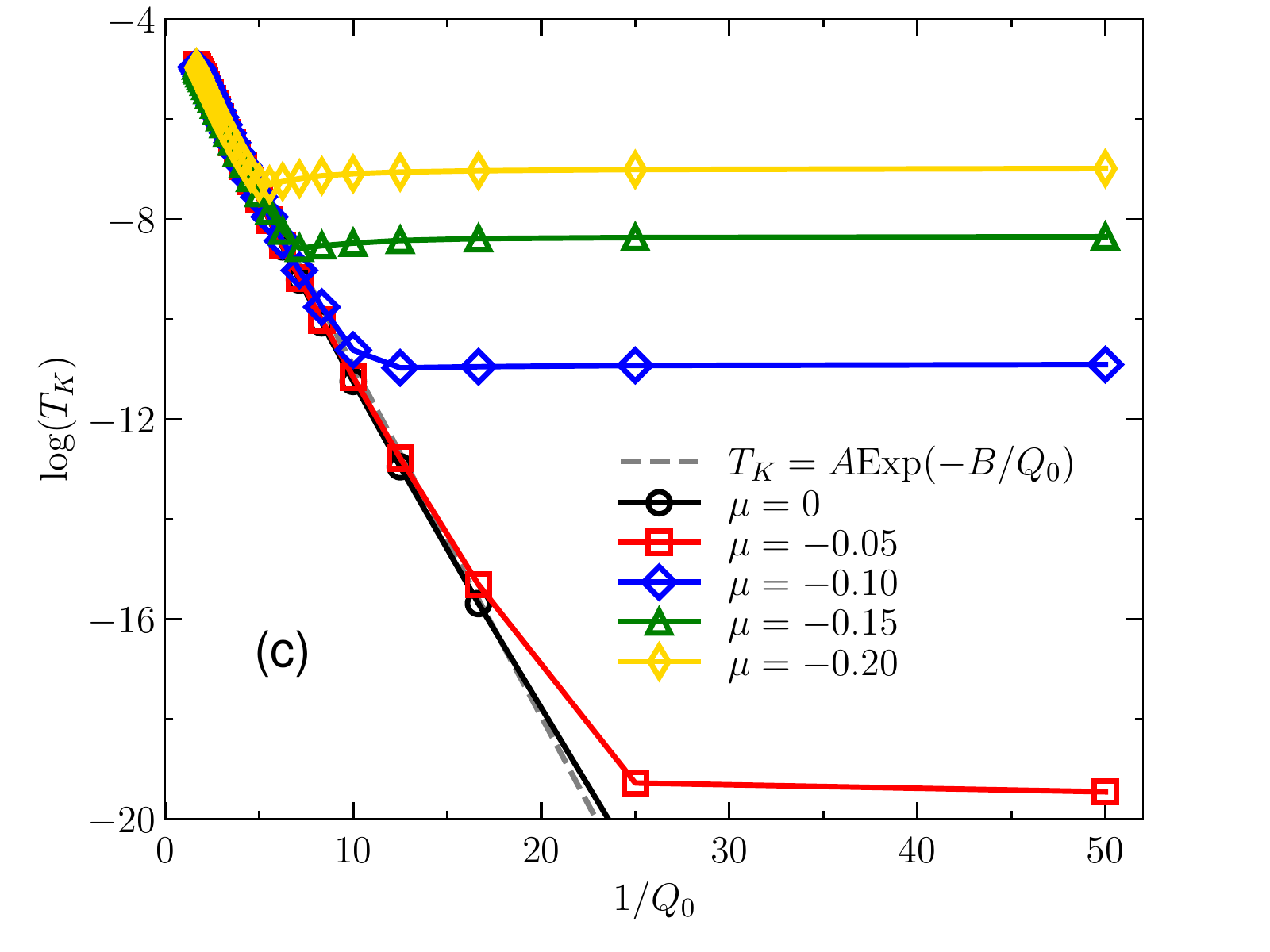}
\caption{
(a) Hybridization function vs energy for $J=2$ and various values of $Q_0$. (b) $T_K(Q_0,\mu)/T_K(Q_0,\mu=0)$ as a function of $\mu/Q_0$, for different values of $Q_0$. (c). Kondo temperature vs $Q_0^{-1}$ for various values of $\mu$. Here, $Q=0$ for all panels.
}
\label{fig:T_k_mu_J=2}
\end{figure}

The situation is more compelling for $J>1$, in which case the energy dependence of the hybridization function changes at $\omega=\pm Q_0$ (for $\mu=0$)~\cite{hybridization_note}.  Figure \ref{fig:T_k_mu_J=2}(a) shows how the hybridization function evolves with increasing $Q_0$ for $J=2$ by fixing $\mu=0$. Clearly, the hybridization function exhibits a plateau of width $2Q_{0}$, consistent with previous analytical calculations in Ref.~\cite{PhysRevB.99.115109}.  Moreover, the height of the plateau increases with $Q_0$, which suggests that the Kondo temperature increases with $Q_0$, resembling the flat band Anderson model. This is important to understand the behavior of $T_K$ in the regime of $Q_0> |\mu|$. Indeed, this suggests that for $J=2$ and for a fixed $Q_0\neq 0$ the system will exhibit two distinct regimes observed in the behavior of $T_K$ as a function of $\mu$.
To confirm this expectation, in  Fig.~\ref{fig:T_k_mu_J=2}(b) we show $T_K(Q_0,\mu)/T_K(Q_0,\mu=0)$ as a function of $|\mu|/Q_{0}$ for several values of $Q_{0}$. One can clearly see that $T_K$ have distinct behavior for $|\mu|/Q_{0}<1$ and $|\mu|/Q_{0}>1$. For $|\mu|/Q_{0}<1$ the Kondo temperature tends to collapse onto a single universal function of $|\mu|/Q_{0}$. On the other hand, for $|\mu|/Q_{0}>1$, the curves change their concavity and are far apart from each other.   
This reveals that the region $|\mu|/Q_0<1$ characterizes a regime dominated by $Q_{0}$ in which $T_K$ should depend very weakly on $\mu$ as the hybridization function is nearly flat. This becomes more evident in Fig.~\ref{fig:T_k_mu_J=2}(c) where we show $T_K(Q_0,\mu)$ vs $1/Q_0$ for various values of $\mu$. These curves expose very clearly that indeed for $Q_0>|\mu|$, $T_K$ depends on $Q_0$ as  $T_{K}=A {\rm Exp}(-B/Q_{0})$. Again, here $A$ and $B$ are quantities dependent on the other parameters of the model. 

\begin{figure}[!t]
\includegraphics[scale=0.5]{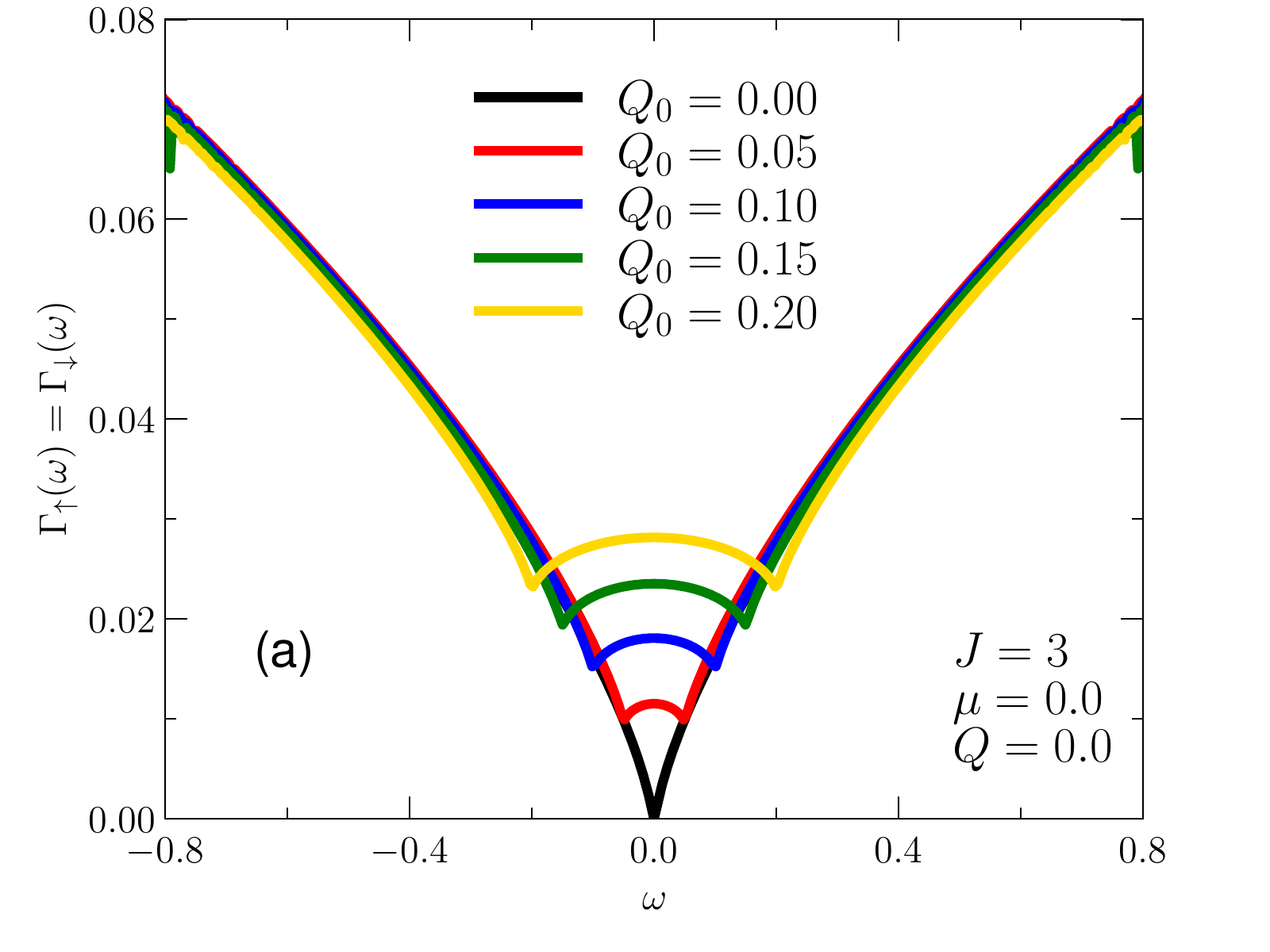}
\includegraphics[scale=0.5]{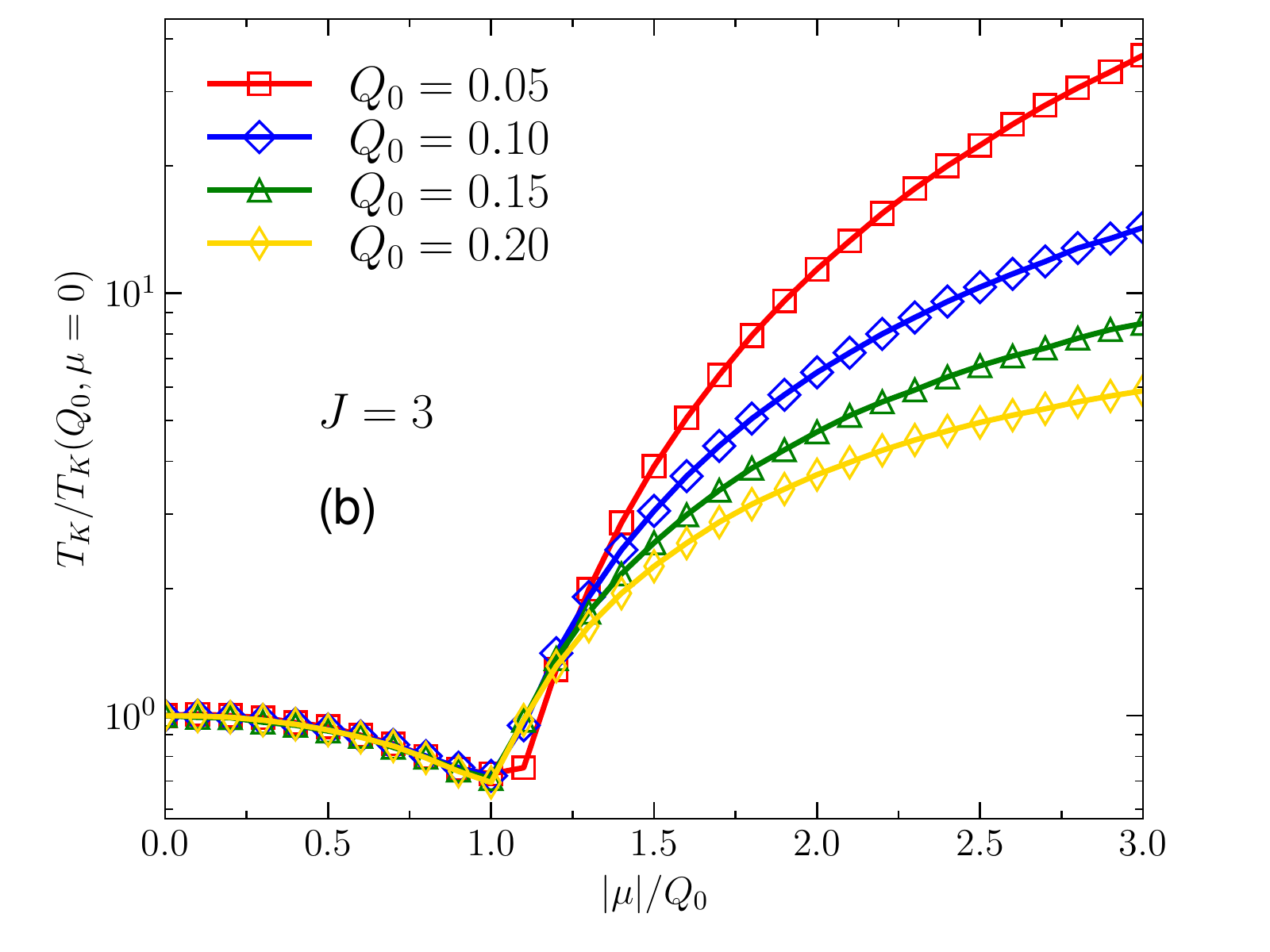}
\includegraphics[scale=0.5]{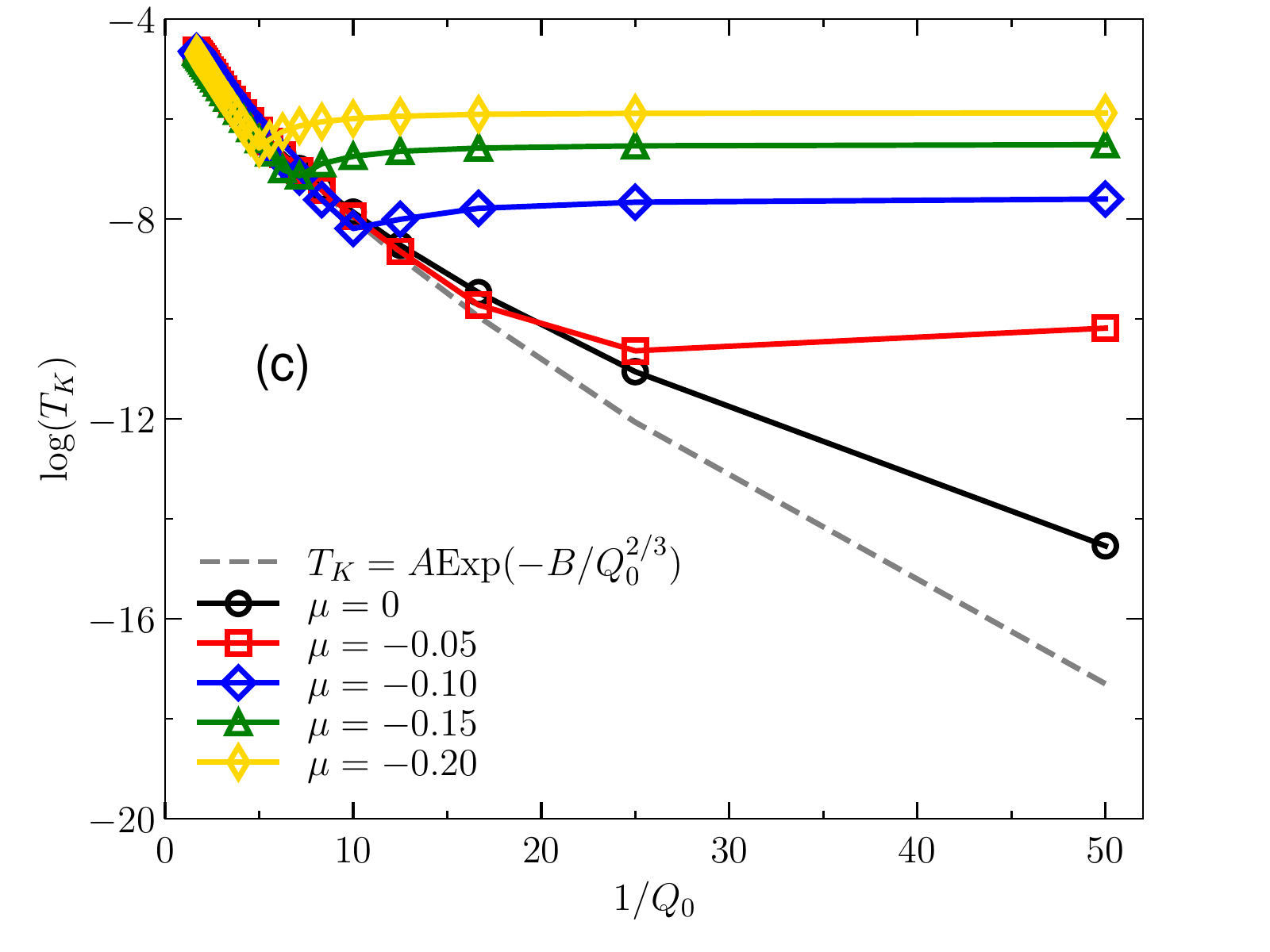}
\caption{
(a) Hybridization function vs energy for $J=3$ and various values of $Q_0$. (b) $T_K(Q_0,\mu)/T_K(Q_0,\mu=0)$ as a function of $\mu/Q_0$, for different values of $Q_0$. (c). Kondo temperature vs $Q_0^{-1}$ for various values of $\mu$. Here, $Q=0$ for all panels.}
\label{fig:T_k_mu_J=3}
\end{figure}

Finally, we examine the Kondo regimes for triple WSM, $J=3$. In Fig.~\ref{fig:T_k_mu_J=3} we show the same results as in Figs.~\ref{fig:T_k_mu_J=1} and \ref{fig:T_k_mu_J=2} but for $J=3$. First, note that we have again two very distinct regions in the energy axis. The flat plateau observed in the hybridization function shown in Fig.~\ref{fig:T_k_mu_J=2}(a) is deformed into an arc within the region $|\omega|\leq Q_0$ in Fig.~\ref{fig:T_k_mu_J=3}(a) with a maximum value at $\omega=0$. The non monotonic behavior of  $\Gamma(\omega)$ is reflected in the Kondo temperature of the system. Figure~\ref{fig:T_k_mu_J=3}(b) shows $T_K(Q_0,\mu)/T_K(Q_0,\mu=0)$ for various valued of $Q_0$. Similar to the case of $J=2$, one can clearly distinguish two regimes separated by $|\mu|=Q_0$. Interestingly, note that the minimum of $\Gamma(\omega)$ observed at $\omega=\pm Q_0$ is accompanied by a minimum of $T_K$ at $|\mu|=Q_0$, visible in green and yellow curves. Moreover, it is remarkable that even in this case the curves collapse nicely onto each other $|\mu|<Q_0$, showing the universality of the $Q_0$ dominated regime. 

From what we have seen above,  the Kondo temperature for IS broken MWSMs behaves asymptotically as $T_{K}=A{\rm Exp}(-B/Q_{0}^{2/J})$ for large  $Q_{0}$ for all $J=1, 2, 3$ [Figs.~\ref{fig:T_k_mu_J=1}(c), \ref{fig:T_k_mu_J=2}(c), and \ref{fig:T_k_mu_J=3}(c)]. This behavior results from the shape of the  effective hybridization function. Recalling that the Kondo temperature depends essentially on the structure of $\Gamma(\omega)$ at $\omega$ near the Fermi level and that $\Gamma(\omega) = \pi V^2\rho_{\rm host}(\omega)$, the behavior of $T_K$ for $Q_0\gtrsim \mu$ can be understood in light of the  well-known Haldane formula for the  Kondo temperature $T_{K}\sim {\rm Exp}[-\pi U/8\Gamma(0) ]$ \cite{Haldane}, valid for a flat band SIAM in particle-hole symmetry point. Keeping this in mind, we can see that $J=2$ is  a special case in which the $\Gamma(\omega)$ becomes flat for $|\omega|\leq Q_{0}$ [Fig.~\ref{fig:T_k_mu_J=2}(a)]. This is why the curves of Fig.~\ref{fig:T_k_mu_J=2}(c) collapse more nicely onto the analytical  result shown in the dashed line as compared to those of Fig.~\ref{fig:T_k_mu_J=1}(c) and Fig.~\ref{fig:T_k_mu_J=3}(c).

\section{Concluding remarks} \label{conclusions}
\label{sec:conclusions}
We have studied the Kondo physics in a quantum magnetic impurity embedded in multi-Dirac(Weyl) node fermionic systems. Our numerical results reveal that the Kondo physics in the double- and triple-Dirac node systems always lie within known classes of pseudogap Kondo problem, well studied in the literature. While no Kondo screening is observed in the particle symmetric case, in the asymmetric case for finite chemical potential, $\mu \neq 0$, the Kondo screening takes place due to finite hybridization function at the Fermi level. However, different scenarios appear in  multi-Weyl node systems. (i) Breaking TRS  (finite $Q$) is detrimental to the Kondo at any finite chemical potential, but the Kondo peak is very slowly suppressed as $|Q|$ increases. This is because the spin polarization induced in the impurity is very tiny, as the polarization of the various conduction and valence band compensate each other, rendering a vanishingly small spin splitting in the hybridization function. (ii) When IS is broken we show that Kondo screening is present for any finite $Q_0$, meaning that the Kondo temperature is always finite.  We find that the double- ($J=2$) and triple- ($J=3$) Weyl node fermionic systems are radically distinct from the single- ($J=1$) Weyl node fermion systems. In contrast to the $J=1$ case where $\mu$ is always an important parameter, for $J=2$ and $J=3$ we observe two distinct regimes: namely  the  regime $|\mu|>Q_0$, in which the Kondo temperature depends strongly on $\mu$, and the regime $|\mu| < Q_0$, where $T_K$ depends very weakly on the chemical potential. In particular, for $J=2$ the system behaves almost as the traditional flat-band single impurity Anderson model. We believe our results expose  distinct Kondo regimes present in multi-node Dirac and Weyl materials, contributing for future theoretical as well as experimental investigations in various Dirac/Weyl available materials.
%

\acknowledgements
We thank Prof. G. B. Martins and Prof. G. S. Diniz for fruitful discussions. The authors acknowledge financial support from CAPES, FAPEMIG. E. V. acknowledges financial support  CNPq, process 305738/2018-6.


\begin{thebibliography}{64}
\bibitem [{kc_()}]{kc_note}%
  \BibitemOpen
  \href@noop {} {}\bibinfo {note} {{We choose the cutoff $k_c$ such that the
  total density of states of the host electrons, normalized within $[-D,D]$ and
  calculated for $Q_0=Q=0$, does not change significantly upon a further
  increase of $k_c$. Note that $D$ is an ultraviolet energy cutoff (imposed
  independently from $k_c$) that establishes a finite bandwidth for the host
  material. }}\BibitemShut {NoStop}%
\bibitem [{Tk_()}]{Tk_note}%
  \BibitemOpen
  \href@noop {} {}\bibinfo {note} {{The Kondo temperature is extracted from the
  magnetic moment, following Wilson criteria $k_BT_K\chi_{\rm
  imp}(T_K)=0.07$~\cite{PhysRevB.21.1003} }}\BibitemShut {NoStop}%
\bibitem [{hyb()}]{hybridization_note}%
  \BibitemOpen
  \href@noop {} {}\bibinfo {note} {{It has been found in
  Ref.~\cite{PhysRevB.99.115109} that the host density of states behaves as
  $\rho_{\rm host}(\omega)\sim |(\omega+\mu)^2+Q_0^2$ for $J=1$ and and
  $\rho_{\rm host}(\omega)\sim |\omega+\mu+Q_0|^{2/J}+ |\omega+\mu-Q_0|^{2/J}$
  for $J=2,3$ }}\BibitemShut {NoStop}%
\bibitem [{KM_()}]{KM_note}%
  \BibitemOpen
  \href@noop {} {}\bibinfo {note} {{An effective equivalent Kondo model could
  be obtained by performing a Schrieffer-Wolff transformation to a single
  impurity Anderson model with parameters equivalent to ours is some particular
  limit }}\BibitemShut {NoStop}%
\end{thebibliography}

%

\end{document}